\documentstyle[epsf]{mn}

\def\figdir{.}

\def\epsscale#1{\epsfxsize=#1\columnwidth}
\def\plotone#1{\par\centerline{\epsfbox{#1}}}

\def\hii{H\,{\sevensize II}}

\def\eg{{e.g.\ }}
\def\etal{{ et al.~\/}}
\def\ie{{i.e.~\/}}
\def\mf{{Minkowski functionals }}
\def\dt{{\Delta T }}

\title{Morphology of the Secondary CMB Anisotropies: the Key to
``Smoldering'' Reionization}
\author[Gnedin \& Shandarin]
{Nickolay Y.\ Gnedin$^1$ and Sergei F.\ Shandarin$^2$ \\
$^1$Center for Astrophysics and Space Astronomy, 
University of Colorado, Boulder, CO 80309, USA; gnedin@casa.colorado.edu\\
$^2$Department of Physics and Astronomy, University of Kansas, Lawrence, KS 
66045, USA; sergei@ku.edu}
\date{}
\pagerange{\pageref{firstpage}--\pageref{lastpage}}

\begin{document}

\def\dim#1{\mbox{\,#1}}

\label{firstpage}

\maketitle

\begin{abstract}
We show how the morphological analysis of the maps of the secondary CMB
anisotropies can detect an extended period of ``smoldering'' reionization,
during which the universe remains partially
ionized. Neither radio observations of the redshifted 21cm line nor IR
observations of the redshifted Lyman-alpha forest will be able to detect
such a period.
The most sensitive to this kind of non-gaussianity parameters are
the number of regions in the excursion set, $N_{cl}$, the perimeter
of the excursion set, $P_g$, and the genus (i.e.\ '1 - number of holes')
of the largest (by area) region. For example, if the universe reionized
fully at $z=6$, but maintained about $1/3$ mean ionized fraction since
$z=20$,then a $2$ arcmin map with $500^2$ pixel resolution and a 
signal-to-noise ratio $S/N=1/2$
allows to detect the non-gaussianity due to reionization with
better
than 99\% confidence level. 
\end{abstract}

\begin{keywords}
cosmic microwave background - cosmology: theory - cosmology: 
large-scale structure of universe -
galaxies: formation - galaxies: intergalactic medium
\end{keywords}

\section{Introduction}

Recent observations of high redshift quasars \cite{bec01,djo01} offer a unique
probe of the physical conditions in the intergalactic medium (IGM) shortly
after the epoch of reionization. Similar future observations will increase
the 
amount of observational data multi-fold and will provide critical
constraints on the theories of galaxy formation.

However, if we want to go beyond the epoch of reionization and to study the
pre-overlap stage during which ionized \hii\ regions expand into still
neutral low density gas, we need to use different bands of electromagnetic
spectrum. Next Generation Space Telescope (NGST) will provide valuable
clues on the earliest episodes of galaxy formation from infra-red
observations, although a relationship between the first galaxies and the
properties of the IGM at high redshifts will not necessarily be easy to
determine from such observations. Radio observations of the redshifted 21\,cm
line of neutral 
hydrogen might be capable of measuring the pre-reionization signal
\cite{mmr97,toz00,ili02,cgo02}, although such a measurement will be at the very
edge of the capability of the next generation radio instruments such as
Low Frequency Array (LOFAR).

Another possible channel to probe the pre-overlap stage of reionization is
studying secondary CMB anisotropies
\cite{hw96,agh96,ksd98,gh98,jk98,pj98,hk99,h00,bru00,das00,ref00,swh01,sbp01,gj01}.
While such a measurement is also in the future, observations of the
secondary CMB anisotropies can provide constraints on the 
physical conditions in the IGM that
are not accessible by other means. In fact, because CMB anisotropies are
sensitive to the total Thompson optical depth, they can probe low levels of
ionization in the mostly neutral gas - which cannot be done by radio
observations of the redshifted 21\,cm line.

Several scenarios have been proposed recently in which the universe
undergoes a protracted episode of incomplete ionization prior to full
reionization, either due to early supernova-driven winds \cite{mfr01,oea01}
or ionizations by energetic X-rays \cite{oh01,vgs01}. This kind of
``smoldering'' reionization (which results in a partial reduction in the
neutral hydrogen fraction) will be virtually impossible to detect with
radio observations. 
Such a signature also cannot be detected by measuring just the spectrum of
the fluctuations, because, as was shown in Gnedin \& Jaffe \shortcite{gj01},
the power 
spectrum of secondary anisotropies is only weakly dependent on the
redshift of reionization. However, Gnedin \& Jaffe \shortcite{gj01} noticed
that the fluctuations themselves were highly non-gaussian. In this paper we
show how tests of the non-gaussianity of the secondary CMB anisotropies can
detect the signature of an epoch of ``smoldering reionization''.

The physical reason behind such a possibility is simple: the earlier the
epoch of ``smoldering reionization'' begins, the more nonlinear the objects
that are responsible for the production of ionizing photons should be, and,
therefore, the more non-gaussian the secondary CMB anisotropies will
be. And it does not matter for this test what the sources of ionizations
really are, only that they form early and, therefore, highly nonlinear at
early times, when the linear fluctuations are smaller.

A big advantage of a morphological approach is that it is virtually
independent of assumptions about the underlying cosmological model. For
example, the redshifted 21cm emission and absorption will depend on the rate
at which ionization front expand into the low density IGM, gobbling up the
neutral gas on the way. And because secondary CMB anisotropies are dominated by
nonlinear structures on small scales \cite{gj01}, they are insensitive to
the specific details of how ionization fronts expand.

\section{Method}

\subsection{Simulations}

In this paper we use the synthetic maps of the secondary CMB anisotropies
from Gnedin \& Jaffe \shortcite{gj01}. These maps were obtained from a numerical
simulation of cosmological reionization, published in Gnedin \shortcite{me1}.

The simulation included all the physical ingredients required to study the
process of reionization, including the 3D radiative transfer. It assumed a
currently fashionable CDM+$\Lambda$ cosmological 
model with $\Omega_{m,0}=0.3$,
$\Omega_{\Lambda,0}=0.7$, 
$h=0.7$, $\Omega_{b,0}=0.04$, $n=1$, $\sigma_8=0.91$ in a 
comoving box with the size of $4h^{-1}\dim{Mpc}$. However, for the purpose
of this paper the specific details about a cosmological model are
unimportant.

For the considered cosmology the angular size of a fixed comoving length is
almost independent of redshift for $z\sim10$. For example, our
$4h^{-1}\dim{Mpc}$ box subtends 2.7 arcmin at $z=5$, 2.2 arcmin at $z=10$,
and 1.9 arcmin at $z=20$. Due to computational limitations, our highest
resolution images include $512^2$ pixels. Thus, we are able to access
multipoles from $l\sim10^4$ to $l\sim5\times10^6$.

Because our goal is to measure the degree of the non-gaussianity of the CMB
maps, we need an appropriate tool for such a task. One of the most powerful
mathematical tools to study different morphologies are Minkowski
functionals.

\subsection{Minkowski Functionals}

Morphology, i.e.\ geometry and topology, of a random field can be quantified
by a set of measures know as Minkowski functionals  \cite{min03}.
Although being known in mathematics since the beginning of 20th
century, they were brought into cosmology not long ago by  
Mecke, Buchert \& Wagner \shortcite{mec-buc-wag94} despite occasional
partial rediscoveries (\eg Gott \etal 1990). In the simplest form \mf 
of a connected region represent
three numbers characterizing the shape of a region: the area $A$, perimeter
$P$, and genus $G$, which can be defined as $G=1-n_h$ where $n_h$ is the
number of holes in the region. \mf are additive, therefore they have 
precise meaning
for any set of the regions, and can be easily computed if they are known
for each region in the set. \mf of the excursion set 
often called the global \mf are particularly
popular in cosmology \cite{got-etal90,sch-buc97,win-kos97,sch-gor98,nov-fel-sh99,nov-sch-muk00,wu-etal01,sh-etal02}. Global \mf of the two-dimensional gaussian 
field is known  analytically \cite{lon-hig57}
\begin{eqnarray}
A(u)&=&\frac{1}{2} \left[1-{\rm erf}\left(\frac{u}{\sqrt{2} \sigma}\right)\right], \cr
P(u)&=&\frac{1}{2R}\exp\left(-\frac{u^2}{2 \sigma^2}\right),\cr
G(u)&=&\frac{1}{(2\pi)^\frac{3}{2}} \frac{1}{R^2}u 
\exp\left(-\frac{u^2}{2 \sigma^2}\right), \label{GMF_G}
\end{eqnarray}
where $R=\sqrt{2}\sigma/\sigma_1$ is the scale of the field;
$\sigma$ and $\sigma_1$ are the rms of the field and its first derivatives 
(in statistically isotropic fields both derivatives 
$\partial u/\partial x$ and $\partial u/\partial y$ have equal rms). 
It is assumed that  $<u>=0$. Here 
$A(u)$ is the fraction of the total area in the
regions of the excursion set, $P(u)$ is the total length of the contour
or total perimeter per characteristic area $R^2$, 
and $G(u)$ is the (number of regions) - (number of holes) per $R^2$.

The popularity of the global \mf is probably related to the existence
of the analytic formulae (eq.~\ref{GMF_G}) and computational easiness.
Computing global \mf is usually done simply
by identifying the pixels satisfying the threshold condition and
applying the Crofton formulae \cite{cro68}. Unfortunately, this
simple technique is quite crude resulting in nonconvergence of
the contour length. For large maps a statistical correction can be
applied \cite{win-kos97}, however its effects in finite size maps is
not quantified. The least effect would be a significant growth of noise.
We use a different technique which identifies every
region in the excursion set and computes its area, perimeter and genus
from the contour points obtained by the interpolation between pixels.
One can find the details of the technique in \cite{sh-etal02}.
In addition to significant increase of the accuracy 
the method allows to use additional information about individual
regions of the excursion set. In this study along with global \mf 
we also use the number of regions (or - using the jargon of the cluster
analysis - number of clusters $N_{cl}$) and 
the \mf of 
the largest area region ($A_p, P_p$ and $G_p$)
labeled as percolating cluster since it
plays a crucial role in identifying the percolation transition. 
The \mf of the percolating cluster of gaussian fields are not not known 
in analytic form but can be easily computed 
(see \eg Shandarin \shortcite{sh02}).
We parameterize
the measurements by the total area in the excursion set $A_g$ measured
in units of the total area of the map. 
The advantages of this parameterization are described in Shandarin 
\shortcite{sh-etal02}.

Since Minkowski functionals depend on the power spectrum ($R$ and $\sigma$
are obviously determined by the power spectrum)
of the
statistical field, we need to separate this dependence from the dependence
on the non-gaussian statistics. For this purpose for each simulated field
of CMB anisotropies we produce 100 gaussian realizations with exactly the
same power spectrum, and analyze these gaussian realization in exactly the
same manner as the simulated field. 
This also reduces the discreteness and boundary effects.
We can then compare Minkowski
functionals for the simulated image and for the gaussian realizations in
order to measure the non-gaussian signal.

\begin{figure}
\epsscale{1.0}
\plotone{\figdir/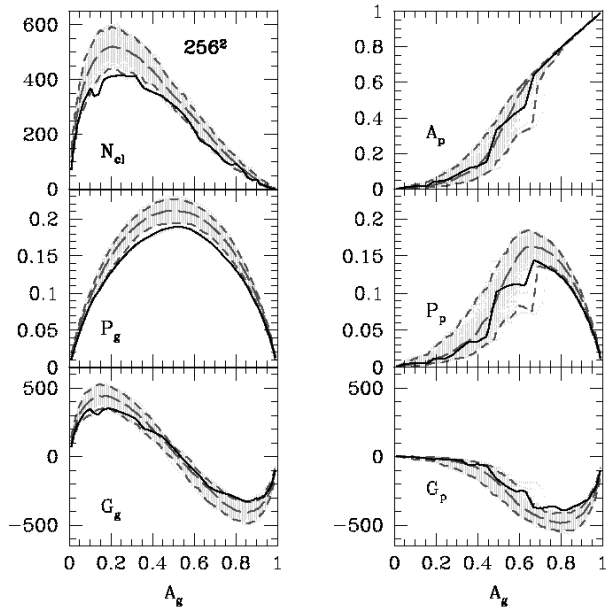}
\caption{\label{figMF} Five Minkowski functionals and the number
  of regions in the excursion set as a function of fractional
  area for the simulated data (black solid lines) and 100 gaussian
  realizations (light grey dotted lines) for a $256^2$ image. 
  Medium grey long and short dashed
  lines show the median and 95\% 
  probability interval for the set of gaussian curves.
}
\end{figure}
An example of such a comparison for all 6 functionals considered here is
given in Figure \ref{figMF}. The behavior of all functionals 
is typical for all grids. 

\begin{figure}
\epsscale{1.0}
\plotone{\figdir/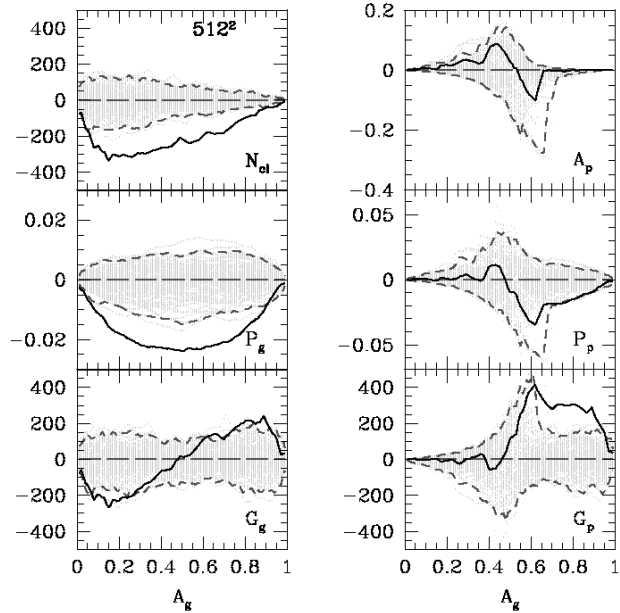}
\caption{\label{figMF512} The deviations of the simulated data 
(black solid lines)
  from the median of 100 gaussian
  realizations (light grey dotted lines) for a $512^2$ image. 
  Medium grey short dashed
  lines show the 95\% 
  probability interval for the set of gaussian curves.
}
\end{figure}

Plotting the deviation from the median of 100 gaussian realizations
better illustrates the type of non-gaussianity in the $\dt/T$ fields
(Figure \ref{figMF512}).
Generally there are fewer regions $N_{cl}$ 
and the global perimeter $P_g$ is shorter than in the corresponding 
gaussian fields. The global genus
$G_g$ at very small $A_g$ (\ie for rare peaks) is close to $N_{cl}$, since
rare peaks have no holes. However, at the other extreme
it is significantly affected by the presence of holes and is 
different from the gaussian genus. The area of the
percolating region does not differ much from the gaussian fields but
the perimeter of the percolating region generally somewhat shorter.
The genus of the percolating region $G_p$ shows more significant deviation
from the gaussian field indicating that the number of holes in 
the largest region 
is smaller than in the gaussian case. Comparing it with the corresponding
part of the global genus (at $A_g > 0.5$) one can see the influence
of small regions: they are present in $G_g$ but omitted in $G_p$.  

One can notice that the number of clusters
$N_{cl}$, the isocontour perimeter $P_g$ and the genus $G_p$ show the
strongest deviations from the gaussian distribution. The global perimeter
$P_g$ is measured  in the grid units and then normalized to one site.

In the rest of this
paper we will focus on $G_p$ and $P_g$ functionals for brevity.

\subsection{Measuring Non-gaussianity}

The task of quantifying the degree of deviation of the simulation from the
gaussian distribution is a formidable one and is generally an unsolved
mathematical problem. Specifically, we would like to know how much gaussian
or non-gaussian are the black curves in Fig.\ \ref{figMF}. Because we
cannot solve this problem in a general case, we have designed the following
simple approach that allows us to measure the degree of non-gaussianity in
some cases and identify cases where this cannot be done.

We start by assuming that we can treat each curve $y(x)$ in the interval
$0<x<1$ as a collection of random variables $y_j=y(x_j)$ where $x_j,
j=1...n$ is a 
subdivision of the interval $[0,1]$ (it does not have to be uniform). We
then treat $y_j$ from gaussian realizations as 
gaussian random variables (this is the main
assumption), so that the joined probability for all $y_j$ for a given curve
$y(x)$ is
\[
  p(y_j)d^ny = \exp\left[-{1\over2}\sum_{i=0}^n\sum_{j=0}^n
    C^{-1}_{ij}(y_i-\bar{y}_i)(y_j-\bar{y}_j)\right],
\]
where $\bar{y}_j\equiv\langle y_j\rangle$ is the mean for $y_j$ and 
\[
  C_{ij} \equiv \langle (y_i-\bar{y}_i)(y_j-\bar{y}_j)\rangle
\]
is the correlation function. Both $\bar{y}_j$ and $C_{ij}$ can be computed
directly from 100 gaussian realizations.

Then for a given collection of $\hat{y}_j$ from the simulated map we can
compute the ``effective'' chi-square
\begin{equation}
   \chi^2_{\rm eff} \equiv \sum_{i=0}^n\sum_{j=0}^n C^{-1}_{ij}(\hat{y}_i-\bar{y}_i)
   (\hat{y}_j-\bar{y}_j).
   \label{ecsdef}
\end{equation}
The problem with this approach is that often $C_{ij}$ is ill-defined if the
curves are smooth and neighbouring points $y_j$ and $y_{j+1}$ are highly
correlated. In order to avoid small eigenvalues in matrix $C_{ij}$ we use
the Singular Value Decomposition to expand the correlation matrix as
\[
   C_{ij} = \sum_{k=0}^n U_{ik}w_kV_{jk},
\]
where $w_k$ are eigenvalues of $C_{ij}$ and matrices $U_{ik}$ and $V_{jk}$
are orthonormal. We can then rewrite the effective
chi-square (\ref{ecsdef}) as a sum over inverse eigenvalues,
\begin{equation}
   \chi^2_{\rm eff} = \sum_{i=0}^n\sum_{j=0}^n \sum_{k=0}^n {1\over w_k}V_{ik}U_{jk}
   (\hat{y}_i-\bar{y}_i)(\hat{y}_j-\bar{y}_j).
   \label{ecssum}
\end{equation}
Let us now consider not the full sum but a partial sum, where we assume
that $w_k$ are sorted in the descending order (they are all positive since the
correlation matrix is always positive definite),
\begin{equation}
   \chi^2_{\rm eff}(l) = \sum_{i=0}^n\sum_{j=0}^n \sum_{k=0}^l {1\over w_k}V_{ik}U_{jk}
   (\hat{y}_i-\bar{y}_i)(\hat{y}_j-\bar{y}_j).
   \label{ecsprt}
\end{equation}
The expression $q(l)\equiv\chi^2_{\rm eff}(l)/l$ as a function of $l$ reaches a
maximum at some value of $l_{\rm max}$. If we interpret $q(l)$ as
chi-square per degree of freedom $l$, then $l_{\rm max}$ gives us the
number of degrees of freedom present in the simulated curve $\hat{y}_j$,
and we choose to truncate the sum (\ref{ecsprt}) at this value of $l$ -
thus, discarding small $w_k$ which makes the inverse correlation matrix
ill-defined. 

Because $q(l_{\rm max})$ is not real chi-square, we cannot use standard
tables to compute the confidence level on our simulated curve. Instead, we
compute $q(l_{\rm max})$ for 100 gaussian curves $y_j$ and compare the
cumulative distribution $P(>q)$ for 100 gaussian realization
with the value $\hat{q}$ for the simulation. If $\hat{q}$ falls within the
range of values that gaussian realizations span, we can directly estimate
the probability that the simulation is non-gaussian by counting how many
gaussian realizations have $q(l_{\rm max})$ greater then
$\hat{q}$. However, because we only have 100 gaussian realization, we
cannot measure probabilities better than 1\% in this way. So, if $\hat{q}$
does not lie within the range spanned by gaussian realizations, we assume
that $q(l_{\rm max})$ for gaussian realizations indeed obey the chi-square
distribution and compute the confidence level on $\hat{q}$ from that
assumption. We also compare the distribution of $q(l_{\rm max})$ with the
true chi-square distribution using the Kolmogorov-Smirnov test, and when
this test fails our assumption becomes invalid. In that case we can only
say that the simulation is non-gaussian at better than 99\% confidence
level.

\section{Results}

\begin{figure}
\epsscale{1.0}
\plotone{\figdir/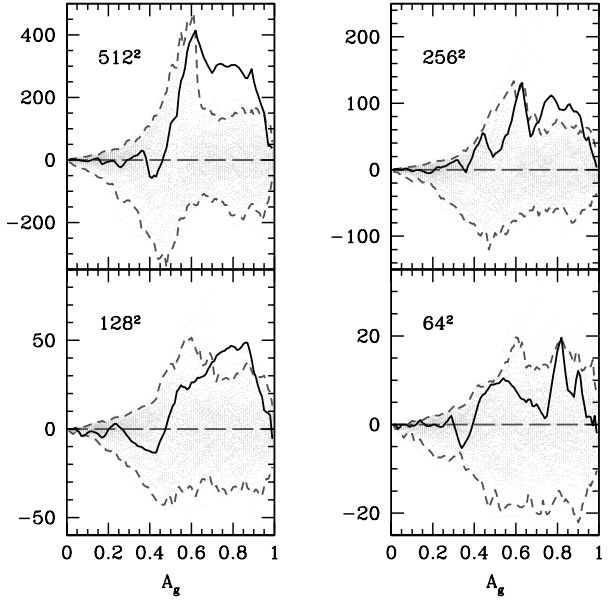}
\caption{\label{figRE} The deviation of the genus of the percolating 
region from the median value of 100 gaussian realizations as a function of 
the total area of the excursion set for four resolutions 
of the simulated image 
  ($64^2$ - 2 arcsec pixel, 
  $128^2$ - 1 arcsec pixel,
  $256^2$ - $1/2$ arcsec pixel, and
  $512^2$ - $1/4$ arcsec pixel). Line markings are as in Fig.\
  \protect{\ref{figMF}}.
}
\end{figure}
Figure \ref{figRE} shows the effect of angular resolution of the
measurement of the Minkowski functionals. For brevity, we only show 
the genus of the percolating region as a function of total area 
of the excursion set for four values of angular resolution
(which also implies different pixel sizes because the image size is fixed
to 2.2 arcmin). In order to provide a better illustration we plot
the deviation of the genus from the median value obtained from 100
gaussian realizations. 
As one can see, higher resolution gives progressively more
non-gaussian signal. 

\begin{figure}
\epsscale{1.0}
\plotone{\figdir/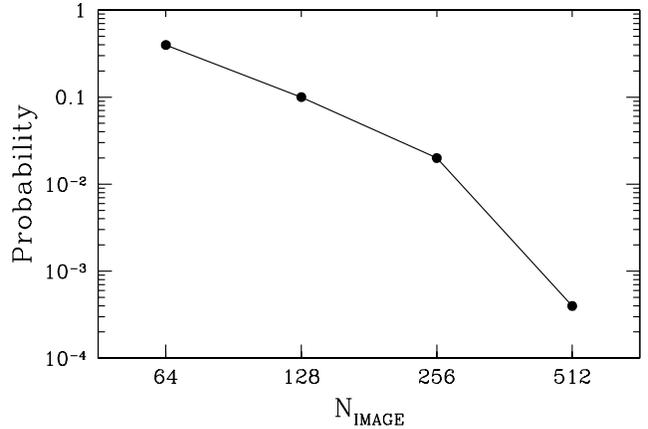}
\caption{\label{figGG} The probability that the simulation image is a
  gaussian realization as a function of resolution (i.e.\ the image size). 
  Higher resolution images are more non-gaussian.
}
\end{figure}
This is illustrated in Figure \ref{figGG}, which shows the probability that
the signal is gaussian as a function of the image size. 
The number of clusters statistic $N_{cl}$ and global perimeter $P_g$
show similar trends.

This trend is easy to understand: as the resolution increases, we probe
progressively smaller spatial scales that are more non-linear - and, thus,
more non-gaussian - than larger scales.

A similar trend is observed when we change the redshift of reionization (we
follow the specific procedure described in Gnedin \&
Jaffe \shortcite{gj01}).
The redshift of reionization is defined as the moment at which the rate of 
change of the mean free path to ionizing radiation peaks. 
It roughly corresponds to the moment when the volume weighted mean fraction of 
neutral hydrogen is about 10\%.

\begin{figure}
\epsscale{1.0}
\plotone{\figdir/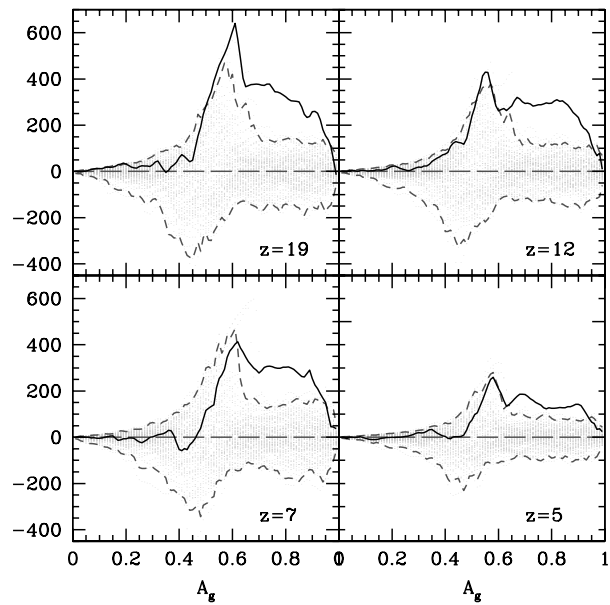}
\hspace{4.1cm} ({\it a\/})
\epsscale{1.0}
\plotone{\figdir/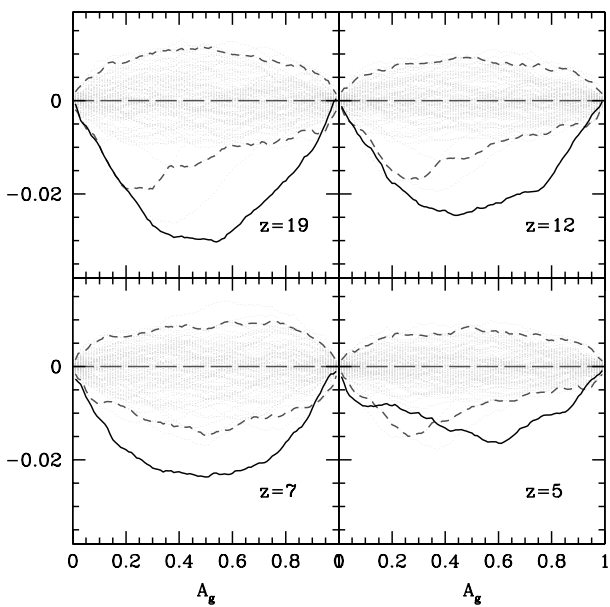}
\hspace{4.1cm} ({\it b\/})
\caption{\label{figEP} ({\it a\/}) deviation from the median value 
of the genus of the percolating
region and ({\it b\/}) deviation from the median value of the perimeter 
of the excursion set as
a function of area for $512^2$ resolution and four different epoch of
reionization: $z=5$, 7, 12, and 19. Line markings are as in Fig.\
  \protect{\ref{figMF}}.
}
\end{figure}

 Figure \ref{figEP} shows the genus and perimeter,
but other Minkowski functionals follow a similar trend. In this figure we 
again show the deviations from the median.

For this result to hold, however, it does not matter whether the universe
is reionized completely or the neutral fraction remains of the order of
10\% or so. After all, the CMB anisotropies are sensitive to the total
optical depth. So, if the mean neutral fraction drops to 10\% at, say,
$z\sim20$, and the complete reionization (during which the neutral fraction
drops to $10^{-4} - 10^{-5}$) takes place at $z\sim6$, our results for
$z=20$ would apply, whereas neither radio observations (because the neutral
fraction is only 10\%) nor optical/IR
observations (because the neutral fraction is still 10\%) 
will be able to obtain any information about the
``smoldering reionization'' period from $z\sim20$ to $z\sim6$.

\begin{figure}
\epsscale{1.0}
\plotone{\figdir/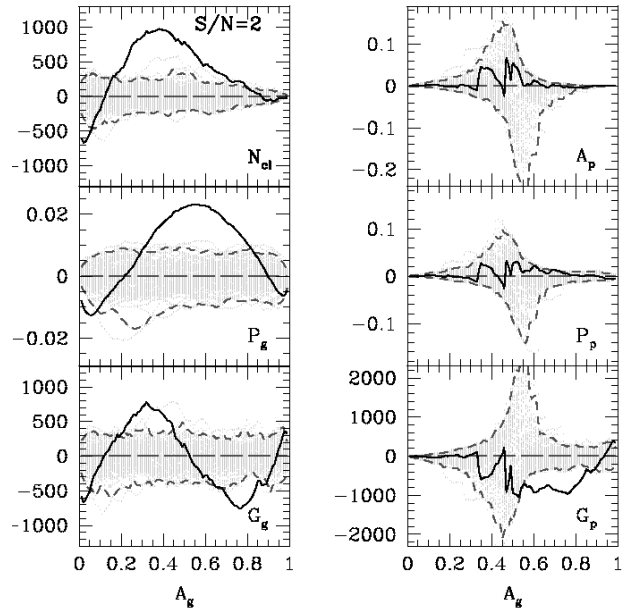}
\caption{\label{figMF512NS05} The deviations of the simulated data 
(black solid lines)
  from the median of 100 gaussian
  realizations (light grey dotted lines) for a $512^2$ image with noise.
  S/N =2
  Medium grey short dashed
  lines show the 95\% 
  probability interval for the set of gaussian curves.
}
\end{figure}

\begin{figure}
\epsscale{1.0}
\plotone{\figdir/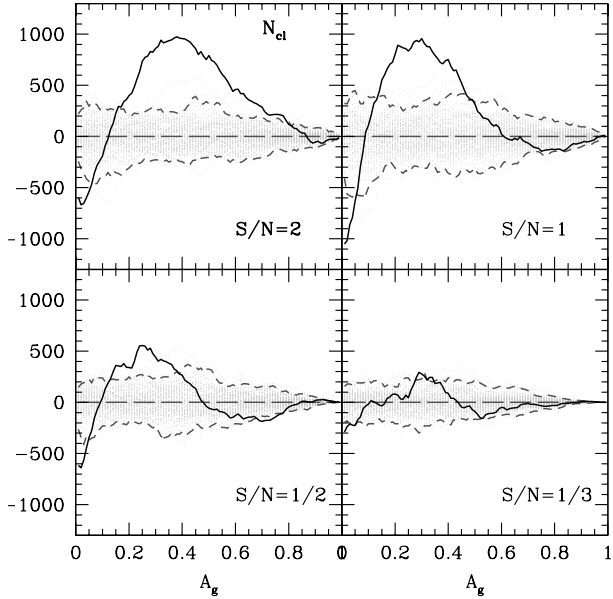}
\caption{\label{figMF512noise} The deviations 
from the median value of the number of regions 
for a $512^2$ image with different ratios of the signal to noise.
Notations are as in fig \ref{figMF512NS05}
}
\end{figure}

Real observations are almost always suffer from noise. 
In order to test the robustness of  the morphological statistic used here, 
we added gaussian noise to the maps of 
the model with  the reionization epoch at $z=12$. 
Besides the model without noise we analyzed the models with the
signal to noise ratios $S/N= 1/5, 1/4, 1/3, 1/2, 1, 2, 3.3, 5, 10, 20, 50$.
We found that increasing noise from zero to about $S/N=2$ the deviations of
the Minkowski functionals from the median of the gaussian ensemble
steadily change from that shown in fig.\ref{figMF512} to 
fig.\ref{figMF512NS05}. The further decrease of the signal to noise
ratio between $S/N=1/2$ and $S/N=1/3$ eventually makes the non-gaussian
signal undetectable as illustrated by fig.\ref{figMF512noise}.
Quantitatively the probabilities of the number
of regions $N_{cl}$ shown in fig.\ref{figMF512noise} to be gaussian
are much less than 1\% for $S/N=2, 1$ and $1/2$ and is about 5\% for
$S/N=1/3$. We remind that making comparison with only one hundred gaussian
fields one cannot reliably estimate this probability much better then
1\%.

This conclusion is also rather insensitive to the precise value of the
neutral fraction during the ``smoldering'' phase. Really, for our values
for the cosmological parameters, the Thompson optical depth accumulated
from the redshift interval from $z_1$ to $z_2$ is
\[
  \tau(z_1,z_2) = c\sigma_T\int_{t_1}^{t_2} n_e dt = 
\]
\[ \phantom{AAA} = 
    2\times10^{-3}\left[(1+z_2)^{3/2}-(1+z_1)^{3/2}\right].
\]
Thus, if we assume that the universe is fully reionized at $z=6$, the
interval from $z=6$ to $z=0$ produces $\tau(0,6) = 0.035$. A period of
``smoldering'' reionization from $z=20$ to $z=6$ with the average free
electron fraction $x_e$ would give another contribution to the Thompson
optical depth of $\tau(6,20)= 0.16x_e$. For example a model in which the
universe reionizes fully at $z=12$ is very similar to the model which
reionizes fully at $z=6$ but has a period of ``smoldering reionization''
from $z=20$ to $z=6$ with $x_e=0.35$.

Thus, until $x_e$ drops below
about $0.035/0.16=0.2$, it will be possible to detect the period of
``smoldering'' reionization. In that case, however, radio observations will
likely be capable of tracing the neutral hydrogen abundance. 

Of course, if the period of ``smoldering'' reionization is shorter, the
sensitivity to the average free electron fraction will be respectively
weaker. However, for the scenario proposed in Venkatesan et
al.\ \shortcite{vgs01}, the effect is indeed measurable.

\section{Conclusions}

We showed how the morphological analysis of the maps of the secondary CMB
anisotropies on sub-arcminute angular scales can detect an extended period
of ``smoldering'' reionization, during which the universe remains partially
ionized. If the neutral hydrogen fraction during such a period is below about
50\% but still well above $10^{-5}$, such a period will be detectable
neither by radio observations of the redshifted 21cm line nor by IR
observations of the Lyman-alpha forest. In that case morphology of the CMB
anisotropies offers the best chance to probe the IGM at early times.

We computed each of six parameters as a function of the fractional area of the
excursion set, $A_g$: 1) the number of regions in the 
excursion set, $N_{cl}$,
2) total perimeter in the excursion set, $P_g$, 3) genus of the
excursion set defined as the number of regions minus the number of holes,
$G_g$, 4) area of the largest (i.e. percolating) region, $A_p$,
5) perimeter of the largest region, $P_p$, and 6) the genus of the
largest region, $G_p$. Three parameters ( $A_g$, $P_g$ and $G_g$) are
known as the global Minkowski functionals of the excursion set, and
$A_p$, $P_p$ and $G_p$ are the Minkowski functionals of largest (by area)
region. We found $N_{cl}$, $P_g$, and $G_p$ are particularly sensitive
to the non-gaussianity in $\dt/T$ maps due to secondary reionization,
they are also the most robust to effects of noise. $A_p$ and $P_p$ are
not sensitive to this type of non-gaussianity, $G_g$ is significantly
less sensitive then $N_{cl}$, $P_g$, and $G_p$. Using $N_{cl}$, $P_g$, 
and $G_p$
one can detect the non-gaussianity in the CMB maps with $S/N=1/2$ at the 
significance
level of better than 99\%.

Morphological analysis of the shapes of individual regions 
in the excursion set provides considerably more information about
non-gaussianity of the maps and potentially may improve the current
result both in terms of the resolution of the maps and signal to 
noise ratio. We reserve this study for the future work. 

\noindent {\bf Acknowledgments:}
S.Sh. acknowledges the support of the GRF 2002 grant at 
the University of Kansas. 
This work was partially supported by National Computational Science
Alliance under grant AST-960015N and utilized the SGI/CRAY Origin 2000 array
at the National Center for Supercomputing Applications (NCSA).

\label{lastpage}

\end{document}